# Structural modifications induced in hydrogenated amorphous Si/Ge multilayers by heat treatments

C. Frigeri[1], M. Serényi[2], A. Csik[3], Z. Erdélyi[3], D. L. Beke[3], and L. Nasi[1]

[1] CNR-IMEM Institute, Parco Area delle Scienze, 37/A, 43010 Parma (Italy)
[2] MTA-MFA Institute, Konkoly-Thege ut 29-33, Budapest H-1121 (Hungary)
[3] Department of Solid State Physics, University of Debrecen, P.O. Box 2, Debrecen H-4010 (Hungary)

*Corresponding author*: C. Frigeri

CNR-IMEM Institute, Parco Area delle Scienze 37/A, Fontanini, 43010 Parma (Italy), frigeri@imem.cnr.it, fax +39 0521 269206

## ABSTRACT

A study is presented of the structural changes occurring as a function of the annealing conditions in hydrogenated amorphous Si/Ge multilayers prepared by sputtering. Annealing changes the structure of the as-deposited multilayer except for the less severe conditions here applied (150 °C, time < 22 h). For higher temperatures and/or times, the modifications consist of layer intermixing and surface degradation in the shape of bumps and craters. They are argued to be due to the formation of H bubbles upon heating. Hydrogen should be mostly released from the amorphous Ge layers.

# 1 - INTRODUCTION

Amorphous multilayers have applications in electronics and optics, e.g., for optical recording and solar cells [1]. Optimization for such applications often requires the selection of appropriate pairs of materials and the creation of solid state solutions by interdiffusion. As a consequence, tailoring of the layer physical properties, like thickness, refractive index, electrical conductivity etc, is achieved. Research is currently underway to measure them in amorphous multilayers because at present only a few data are available. In the case of Si/Ge amorphous multilayers (MLs) the improvement of the physical properties is also accomplished by enriching them with hydrogen because H passivates the dangling bonds of Si and Ge that act otherwise as efficient electronic levels [2, 3]. H stability is of primary importance in order to avoid de-passivation of the dangling bonds. It will be shown here that H behaviour also influences the structural characteristics of the amorphous layers. To optimize the amorphous layers properties one must also control the diffusion of the elements which is quite hard because at present the mechanisms of diffusion are not fully understood. Investigation of the overall thermal stability of these multilayers is thus of great interest for the interpretation of their operation and prediction of their lifetime.

In previous works it was experimentally shown that in amorphous Si/Ge MLs the diffusion is very asymmetric because of the strong concentration dependence of the interdiffusion coefficient [4-6]. This strong concentration dependence leads to a significant initial curvature on the $lnI/I_o$ vs. $t$ curve, as well ($I/I_o$ is the normalized intensity of the first-order small-angle X-Ray diffraction peak, $t$ the time). In agreement with our calculations the silicon could enter into the germanium layer but germanium cold not diffuse into silicon. At the same time the Ge layer became thicker, the initially sharp interface at the start of the process does not flatten but shifts consuming the Si layer. Calculation can show that with increase of the Si concentration the process of diffusion slows down at longer times. It was expected that in hydrogenated Si/Ge MLs hydrogen could affect diffusion of Si. However, it was seen that after a short annealing time significant structural changes occurred at the surface of the hydrogenated Si/Ge MLs. This urged the study of the relationship between annealing conditions and structural modifications in such samples. The results of this study are presented here.

## 2 - EXPERIMENTAL

The amorphous Si/Ge multilayers have been deposited on (100) Si substrates with a conventional radio frequency sputtering apparatus (Leybold Z 400). Before sputtering, the chamber was evacuated to $1 \times 10^{-4}$ Pa . The targets were presputtered for 10 min. A 1.5 kV wall potential was applied to the Si and Ge targets. Argon (purity 99.999 %) was used as sputtering gas. During sputtering, the substrate was water cooled and the chamber pressure was kept at 2.0 Pa. The multilayers consisted of a sequence of 50 couples of Si and Ge layers. Each Si and Ge layer was 3 nm thick as confirmed by Transmission Electron Microscopy (TEM). The hydrogenated Si/Ge samples have been deposited by adding hydrogen in the sputtering chamber at a flow rate of either 3 or 6ml/min. Both not-hydrogenated and hydrogenated multilayers have been submitted to the following heat treatments (HT) in high purity argon (99.999%): a) HT1 at 250 °C for 0.5 h + 450 °C for 5 h, b) HT2 at 350 °C for 1 h, or c) 150 °C for several times + 250 °C for several times.

The samples have been analysed by cross sectional TEM, Scanning Electron Microscopy (SEM), Atomic Force Microscopy (AFM) and Small-Angle X-Ray Diffraction (SAXRD). The TEM specimens were prepared by mechanical thinning of sandwiches, containing a small piece of the sample, down to 30 μm followed by Ar ion beam thinning down to electron transparency. SEM and AFM were used for surface morphological examinations. Θ−2Θ SAXRD was employed to investigate the diffusional processes by measurement of the attenuation of the diffraction peaks [4-6] using CuKα radiation.

## 3 - RESULTS AND DISCUSSION

All the annealed Si/Ge multilayers remained amorphous after the heat treatment as seen by electron diffraction. The not-hydrogenated samples did not exhibit any morphological surface degradation after annealing with respect to the unannealed ones as seen by AFM (Fig. 1): the rms value (0.193 nm) was practically the same for both type of samples.

On the other hand, in the hydrogenated samples submitted to the heat treatments HT1 and HT2 a remarkable change of the surface morphology has be detected. Holes, even as large as 9 μm, formed on the surface which could be as deep as the whole ML thickness as measured by SEM, AFM (Figs. 2-3) and TEM. Besides craters, AFM images also show that the sample surface has bumped up in many places, which suggests that bubbles have formed inside the hydrogenated multilayers. The AFM images of the craters also suggest that they are due to a kind of explosion. The craters density is higher in the samples submitted to the more severe heat treatment (type HT1). All this would lead to the conclusion that during heat treatment gas bubbles formed in the MLs, very likely due to the enhanced gathering of H atoms, that eventually blew up destroying locally the multilayer. By TEM it has been seen that locally there exist areas inside the MLs where a pronounced intermixing of the layers has occurred, sometimes accompanied by small holes at the surface (Fig. 4). Such localized areas should be the first steps of the formation of the hydrogen bubbles hypothetised above. Around such areas of layers intermixing the original ML structure is preserved (Fig. 4), with no escape of H, which explains why the small angle X-ray spectrum exists in the annealed MLs as well, as shown in Fig. 5, and why diffusional intermixing could be measured [4-6].

In order to determine a temperature suitable for diffusion experiments without destroying the structure of the hydrogenated samples the preannealing temperature was reduced to 150 °C (heat treatment HT3). Surface bumps started to appear for annealing times as long as 22 h. No surface deterioration was observed for annealing at the shorter times of 0.25, 0.5, 1 and 8 h. However, even for the latter not degraded samples heat treatments at 250 °C for only 15 min were able to produce surface craters and bumps (Fig. 6) whose density increased with increasing annealing time. Thus, pre-annealing at 150 °C, 22 h is not able to reduce the H concentration to values that allow to anneal at higher temperatures without destroying the surface. The small angle X-ray diffraction spectrum did not change for every combination of the annealing time and temperature of the heat treatment of type HT3 (Fig. 7), which means that no significant diffusion takes place in the samples.

According to literature in crystalline and amorphous Si and Ge the binding energy of the Si-H bond is higher than for the Ge-H bond, with hydrogen attaching at the dangling bonds [7, 8]. It is thus to be expected that the Ge-H bond is less stable when energy is supplied by the heat treatments. Evolution of H from a-Ge:H at annealing temperatures lower than from a-Si:H has been shown [2]. The hydrogen bubbles should therefore mainly be due to release of hydrogen from the Ge layers. Breaking of the Si-H bonds is also possible but should play a minor role. It has been reported that the incorporation of H may take place in such a way that any additional H atom is incorporated at a smaller binding energy than hydrogen atoms already present before, which had reached the substrate earlier [10]. This further increases the probability of formation of H bubbles in the more hydrogenated (6 ml/min) samples, besides the more straightforward reason that they contain more H. Once an interconnect void network is formed H outdiffusion from this network becomes more and more efficient [10].

## 4 - CONCLUSIONS

By studying the structural changes occurring in hydrogenated amorphous Si/Ge multilayers because of heat treatments the following has been observed:

a) for heat treatment HT1, large surface craters and bumps form accompanied by localized layer intermixing. TEM observations show that the original layer structure is preserved outside the intermixing areas.

b) for heat treatment HT2, the same features as for the case a) are observed but the density of surface craters and bumps and of the intermixing areas is smaller.

c) for heat treatment HT3, the samples are featureless for annealing at 150 °C for times smaller than 22 h. They exhibit bumps and craters for 22 h annealing as well by applying a subsequent annealing at 250 °C for any time whatever is the structural conditions of the samples pre-annealed at 150 °C.

The structural changes are due to the formation of hydrogen bubbles in the MLs that eventually blow up by increasing the annealing temperature and time. Hydrogen is first released from the Ge layers because of the lower binding energy of the Ge-H bond. The capability of SAXRD to measure diffusional intermixing of Ge and Si even in samples with highly degraded surface (samples of type HT1 and HT2) is due to the preservation of areas of undeteriorated layer sequence in the ML stacks.

## ACKNOWLEDGEMENTS

This work was supported by OTKA Grant No. D-048594, K61253, 70181, F043372 and by the Scientific Cooperation Agreement between MTA (Hungary) and CNR (Italy). Z. Erdélyi acknowledges support from Bolyai János Foundation.6

**FIGURE CAPTIONS**

Fig. 1 - AFM image of a not hydrogenated sample submitted to heat treatment HT1. RMS = 0.193 nm on the whole sampled area. Lateral scan size 2 μm.

Fig. 2 - SEM image of the surface of a hydrogenated sample (hydrogen flow rate 6 ml/min) submitted to heat treatment HT2.

Fig. 3 - AFM image of the surface of a hydrogenated sample (hydrogen flow rate 6 ml/min) submitted to heat treatment HT1.

Fig. 4 - TEM image of a hydrogenated sample (hydrogen flow rate 6 ml/min) submitted to heat treatment HT2. An area of layer intermixing is visible in the centre of the picture. White arrow indicates where it touches the top surface. Regular layer sequence is visible at top, bottom and left sides of the intermixing area. The ML is here detached from the Si substrate.

Fig. 5 - SAXRD spectrum of a hydrogenated sample (hydrogen flow rate 6 ml/min) submitted to heat treatment HT2.

Fig. 6 - SEM image of the surface of a hydrogenated sample (hydrogen flow rate 6 ml/min) submitted to heat treatment HT3 (see text).

Fig. 7 - SAXRD spectra of a hydrogenated sample (hydrogen flow rate 6 ml/min) submitted to heat treatment HT3. a) As-deposited, b) after annealing at 150 °C for 22 h.

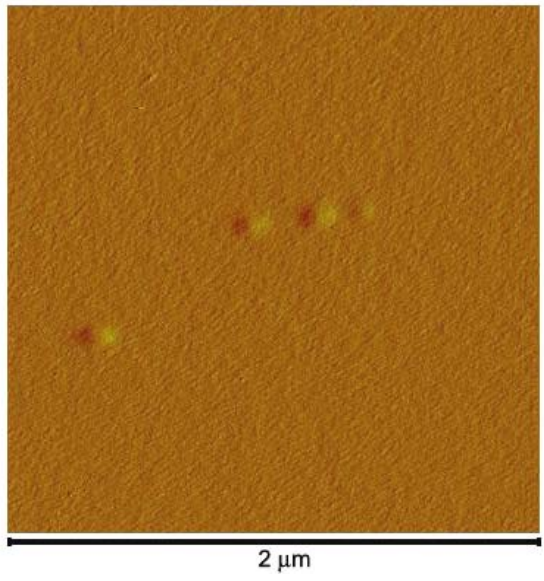

Fig. 1

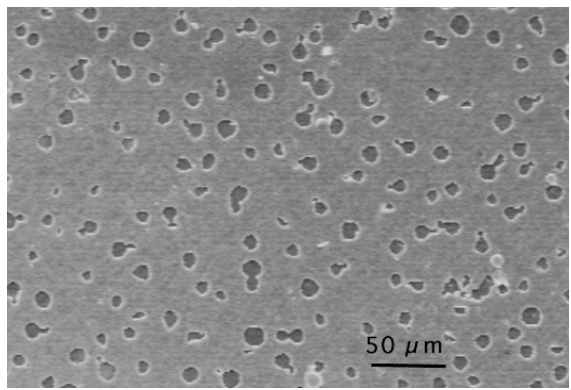

Fig. 2

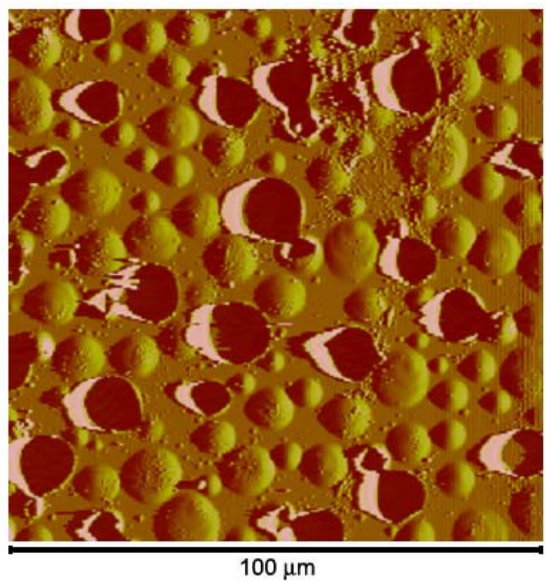

Fig. 3

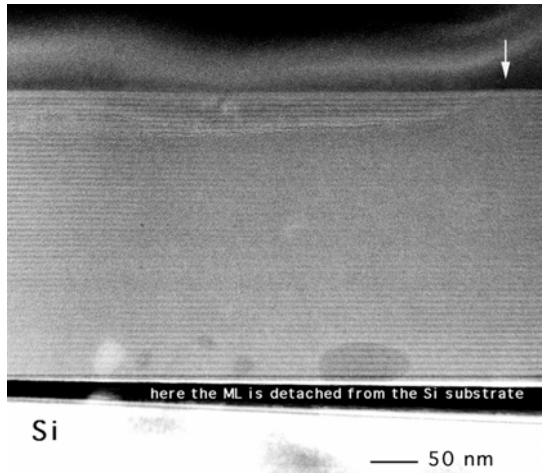

Fig. 4

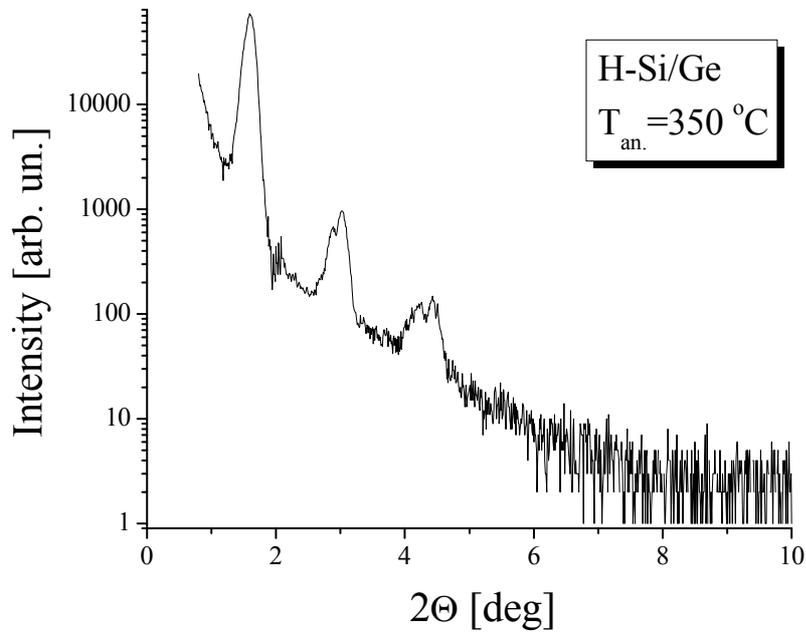

Fig. 5

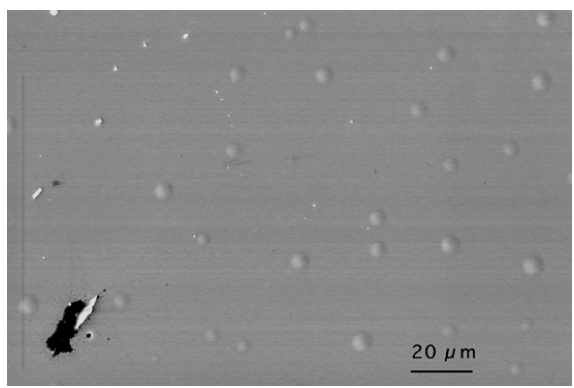

Fig. 6

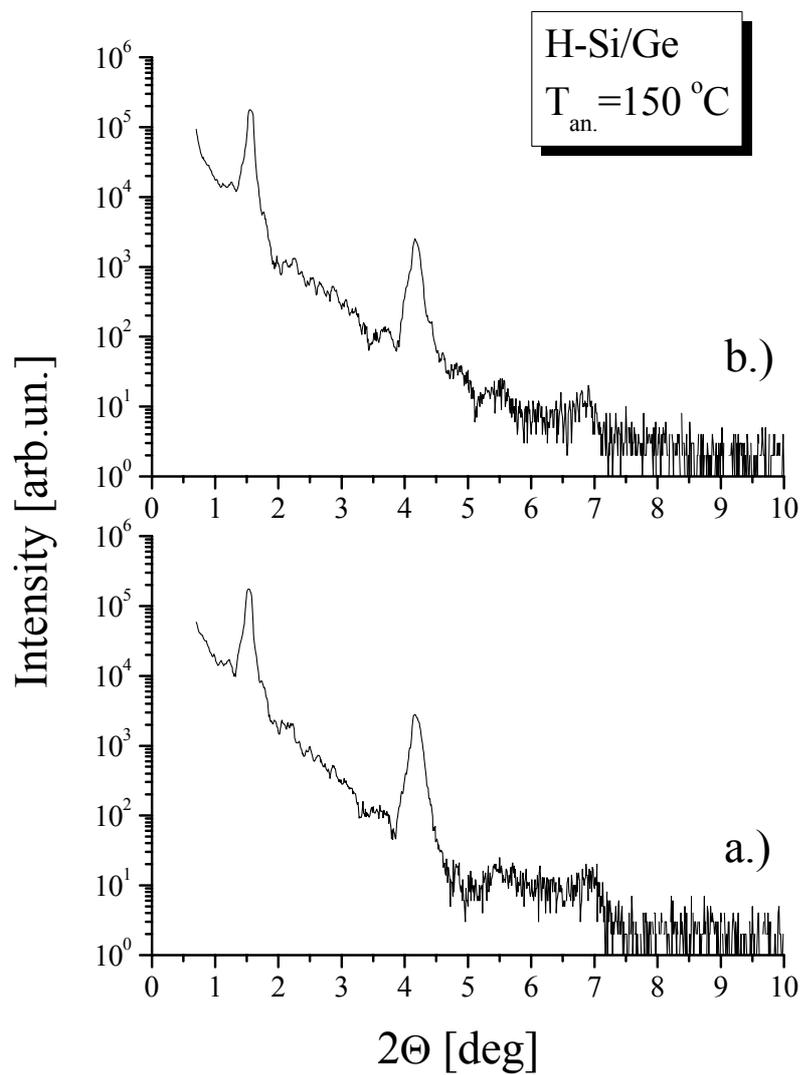

Fig. 7

12